\documentclass[11pt]{article}
\usepackage{geometry}
\usepackage{xcolor}
\usepackage{graphicx}
\usepackage{amsmath}
\usepackage{amssymb}
\usepackage{cite}
\newcommand{\be}{\begin{equation}}
\newcommand{\ee}{\end{equation}}

\newcommand{\Rmnum}[1]{\expandafter\@slowromancap\romannumeral #1@}
\newcommand{\bea}{\begin{eqnarray}}
\newcommand{\eea}{\end{eqnarray}}
\begin{document}
\def\C{{\mathbb{C}}}
\def\R{{\mathbb{R}}}
\def\s{{\mathbb{S}}}
\def\T{{\mathbb{T}}}
\def\Z{{\mathbb{Z}}}
\def\W{{\mathbb{W}}}
\def\Bbb{\mathbb}
\def\BZ{\Bbb Z} \def\BR{\Bbb R}
\def\BW{\Bbb W}
\def\BM{\Bbb M}
\def\BC{\Bbb C} \def\BP{\Bbb P}
\def\CP{\BC\BP}
\begin{titlepage}
\title{Modified gravity in the interior of population II stars} 
\author{} 
\date{
Shaswata Chowdhury, Tapobrata Sarkar 
\thanks{\noindent 
E-mail:~ shaswata, tapo @iitk.ac.in} 
\vskip0.4cm 
{\sl Department of Physics, \\ 
Indian Institute of Technology,\\ 
Kanpur 208016, \\ 
India}} 
\maketitle 
\abstract{We study the effects of a beyond-Horndeski theory of modified gravity
in the interior of a population II star. We consider a simple phenomenological model of a 
$1.1M_{\odot}$ star that has left the main sequence, has a thin Hydrogen burning shell with a partially degenerate
isothermal core, surrounded by a radiative envelope having two regions of distinct opacities. 
Using suitable matching conditions at the two internal boundaries, a numerical analysis of the resulting
stellar equations in modified gravity is carried out. While overall, gravity may be weakened, resulting in a decrease of 
the luminosity and an increase of the radius of the star, these effects are reversed near the core. 
It is suggested how the model, within its limitations, can yield a bound on the modified gravity parameter.}
\end{titlepage}

\section{Introduction}

The theory of general relativity (GR), formulated by Einstein more than a century ago, is the most successful theory 
of gravity, and has been validated by several precision tests. More recently, issues relating to the observed cosmic acceleration 
and the cosmological constant seem to indicate the necessity of modifications to GR, where one might add extra 
degrees of freedom to the Einstein-Hilbert action of GR. Such theories, popularly termed as 
``modified gravity'' (see, e.g. the excellent reviews of \cite{CliftonRev} - \cite{KaseRev}), 
are becoming increasingly popular in the literature. Some of the best studied {\it avatars} of
modified gravity are the so called scalar-tensor theories (STTs). In these scenarios, the compatibility of modifications to GR effects 
in solar system tests, necessitate invoking some kind of screening mechanism, the most efficient being the Vainshtein
mechanism \cite{Vainshtein} (see, e.g. \cite{JainKhoury}, \cite{BabichevRev} for reviews), where GR is recovered in the 
near regime via a non-linear screening of modified gravity. 

The most general physical (i.e ghost-free) theories of a scalar field coupled to gravity constitute the so called 
Horndeski theories \cite{Horndeski}. Recall that, according to Lovelock's theorem, Einstein's equations constitute the 
unique set of second order equations that is obtained from a Lagrangian that depends on the metric, and its first and
second order derivatives. A minimal additional degree of freedom alluded to above, involves a scalar field, called
a Galileon. This was constructed in the work of \cite{Nicolis}, its covariant version and a further generalization 
was derived in \cite{Deffayet1}, \cite{Deffayet2} and the latter was shown \cite{KobayashiH} to be equivalent
to Horndeski theory. Closely related to these are the GLPV theories \cite{BH1},\cite{BH2} which extend the 
generalized Galileon theory to the ``beyond Horndeski'' class, which are physical, i.e., are free from ghost modes.
The degenerate higher order scalar tensor theories beyond Horndeski, in which the GLPV models
arise as special cases arose in the works of \cite{Langlois1} - \cite{Zumalacrregui}. 
This was followed by the important work of Kobayashi, Watanabe and Yamauchi \cite{KWY}, who demonstrated 
that in such beyond-Horndeski theories of gravity, the Vainshtein mechanism might be partially effective, i.e., inside 
a stellar object, the effect of modified gravity is not screened. This was further studied in \cite{Crisostomi}, \cite{Langlois3},
in the wake of the recent discovery of  gravitational waves (GW170817) and stringent restrictions were put on the
allowed theories. 

A remarkable consequence of the partial breaking of the Vainshtein mechanism inside stellar objects results in the
fact that in the low energy (Newtonian) limit, the pressure balance equation 
inside astrophysical objects is modified. For the beyond-Horndeski theories that we consider here, 
the modification of the pressure balance equation occurs via an additive term involving a dimensionless
parameter $\Upsilon$ (defined subsequently in subsection \ref{modgrav})
which represents the effect of the extra degrees of freedom, and renormalizes the 
Newton's constant. Since the pressure balance equation is a crucial ingredient in the derivation of analytical 
formulae corresponding to astrophysical observables that can be experimentally verified, 
one is immediately led to the conclusion that the theoretical 
results of this class of modified gravity theories can be constrained by experimental data. Indeed, a number of 
works in this direction have been reported in the last few years, starting
from the pioneering work of Koyama and Sakstein \cite{SaksteinPRD}, see \cite{Saito} - \cite{Tapo2}. 
For a recent review of the effects of modification of gravity in stellar objects, see \cite{SaksteinRev}, \cite{GonzaloRev}.

Quite naturally, most of the above mentioned works have focussed on cases where observational
signatures can be used to constrain modified gravity in astrophysical scenarios. 
The results obtained therein confirm that the theories of modified gravity 
described above broadly result in a weakening of gravity inside stellar objects if $\Upsilon >0$ and 
a strengthening of gravity if $\Upsilon < 0$  (see, e.g., \cite{SaksteinPRD}, \cite{Sakstein}, \cite{ws1}, \cite{Saltas}). 
This is true at least far from the stellar core, as follows from the
modified pressure balance equation, when the density of the stellar object falls away from the center, 
as is usually the case. It is precisely due to this reason that main sequence stars in modified gravity, for example, 
are ``dimmer and cooler'' \cite{SaksteinPRD} than what one would normally expect from a GR analysis of the same star. 

Not much work however, has been reported on the internal structure of stars in modified gravity. 
In fact, most of the astrophysical studies of the subject till date have been carried out in situations where 
the entire stellar object is either in radiative or convective equilibrium, and can be approximated as a polytrope. 
Then, one usually assumes a suitable polytropic equation of state in 
studying the effects of the modification of the pressure balance equation. The situation however becomes more interesting 
and involved, in the presence of a distinct core-envelope structure, with a thin intermediate Hydrogen burning shell. 
It is well known that such a scenario might be difficult to deal with analytically -- there are a large number of equations 
that one has to handle, even in a simplified model. In this paper, we carry out a detailed analysis of such a case, with a 
phenomenological model proposed long back by Hoyle and Schwarzschild \cite{HoyleSchwarzschild} (HS). 
Our results indicate that while overall the ``dimmer and cooler'' picture of \cite{SaksteinPRD} is indeed true, 
there is rich physics in the stellar interior in the presence of modified gravity. 
In particular, we find that the core radius of the star decreases, and its shell density increases with $\Upsilon$, while the
core temperature decreases -- results which might seem counter-intuitive given that increasing $\Upsilon$ 
weakens gravity, but we will explain why this is true. We also show how possible bounds on
$\Upsilon$ can be given in such models. 

We remind the reader at the outset that we are not constructing evolutionary tracks of stars here, a topic
already studied earlier (there are several publicly available state of the art codes that accomplishes this). 
Neither are we attempting to construct an exact model of a stellar interior in 
modified gravity. Our study will be purely phenomenological, and based on the approximate model that
we consider, and our aim is to obtain an analytic handle on the stellar structure equations in 
modified gravity scenarios. 

This paper is organized as follows. In the next section 2, we will formulate the problem, by first elaborating on 
our model, reviewing the necessary equations in the Newtonian limit of GR, and then set up the modified
equations in beyond Horndeski theories. Section 3 contains a detailed exposition to our numerical analysis 
and results. The paper ends with discussions and conclusions in section 4. 

\section{The Model and the setup}

We now present our model and the modified gravity setup. This section has four parts. We first elaborate upon the
model, and the various approximations used, in section \ref{model}, review the modified gravity setup in \ref{modgrav},
write the stellar equations in the Newtonian formalism in section \ref{steq}, and then derive the ones that will be 
used for our numerical analysis in the modified gravity scenario, in section \ref{Transformed}. 

\subsection{The model}
\label{model}

Following HS, we model a $1.1M_{\odot}$ 
star that has left the main sequence and study the new features of the stellar interior that modified gravity predicts.\footnote{For a sampling of the 
older literature on the topic of evolution of low mass stars like the one considered here, see e.g. \cite{low1}-\cite{low5}.} 
Indeed, a reasonable place to look for effects of modified gravity seems to be metal-poor population II (pop-II) stars in 
globular clusters, which have left the main sequence (population III stars which are even poorer in metals will not be 
of much relevance in the absence of data as of now), as in a phenomenological model, the metallicity can be 
often ignored. 

One has to make a large number of simplifying assumptions to deal with these
stars analytically. These are standard assumptions in the literature (see, e.g. \cite{Schwarzschild}),
and following HS, we will assume that in our model, the metallicity, radiation pressure and degeneracy
arising out of relativistic effects are all negligible. 
The first assumption of zero metallicity needs some explanation. Here, as we will see in 
subsection \ref{steq}, metallicity enters only by the Kramer's formula for the opacity $\kappa$ for 
free-free transitions, namely,
\begin{equation}
\kappa = 3.68\times 10^{22}\left(1-Z\right)\left(1+X\right)\frac{\rho}{T^{3.5}}~.
\label{Kramer}
\end{equation}
The approximation that we use is, $1-Z \simeq 1$. 

Further, it is assumed that the (partially degenerate) core has a constant composition
of Helium. The envelope, which is in radiative equilibrium, also has a constant composition predominantly of Hydrogen. 
Further, the core mass fraction is taken to be $0.10$ times the total mass (we will comment on this shortly).
The thickness of the Hydrogen burning shell between the core and the envelope is assumed to be negligible. 
Further the envelope is assumed to consist of two parts. The opacity of the inner part of the envelope arises due
to free electron scattering and that in the outer part is determined by the free transitions 
of Hydrogen and Helium. Finally, the temperature and pressure of the star is taken to be zero at the surface. 
In our model, we incorporate a further assumption that simplifies the numerical computations. 
Namely, we will assume that the dominant source of nuclear energy in the shell is 
through the carbon cycle (for an elaborate treatment of nuclear energy generation in stellar objects, see, e.g., 
\cite{nucbook}). This will be justified in sequel. 

\begin{figure}[t]
\centering
\includegraphics[width=3.7in,height=2.5in]{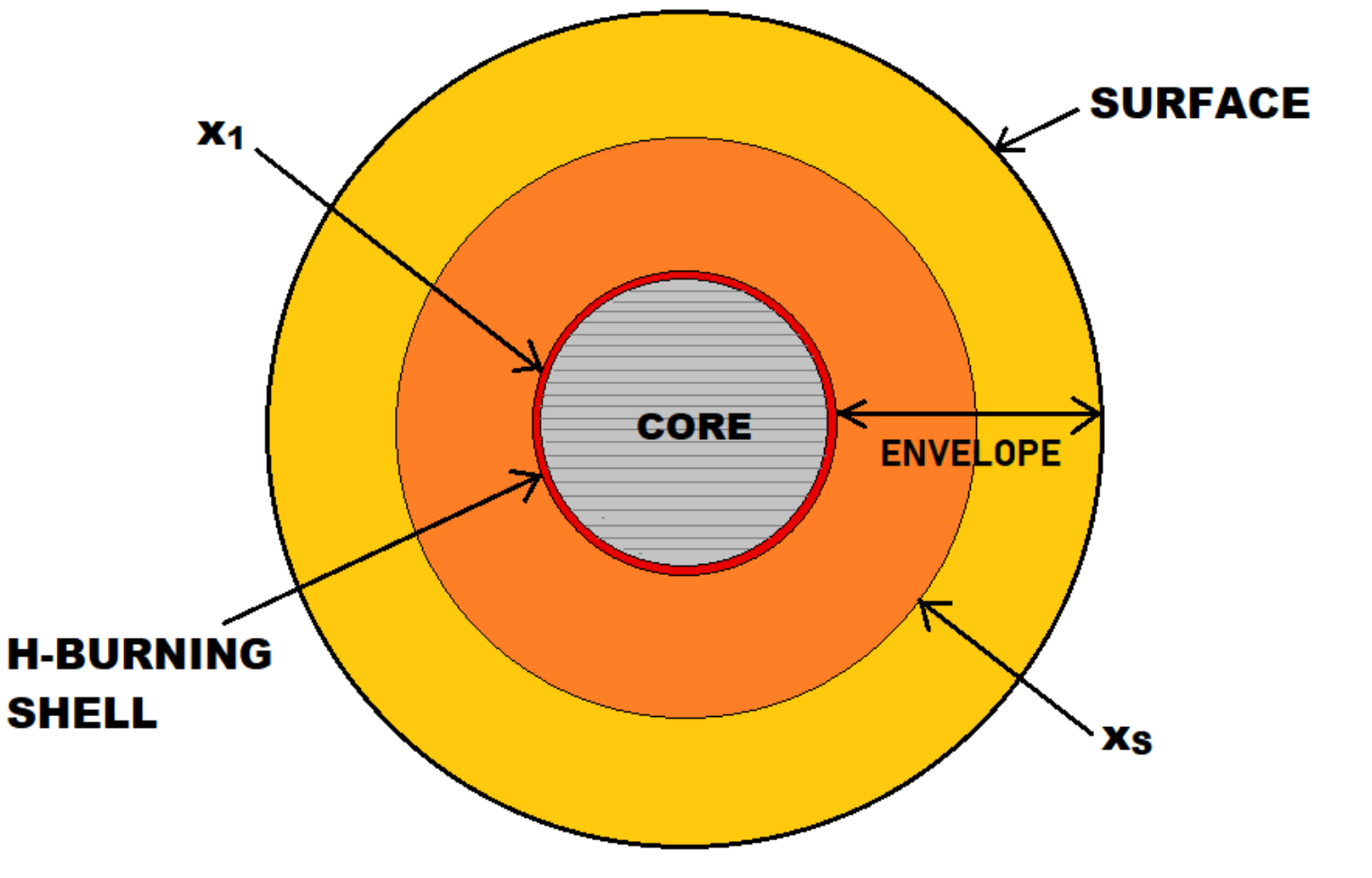}
\caption{Schematic Diagram for the stellar model used in this paper.}
\label{fig2}
\end{figure}
That this is a very good model for a $1.1M_{\odot}$ star near the turnoff point in the Hertzsprung-Russell
diagram was recognized by HS, who also noted
that in this particular setup, a higher mass (say $M=1.2M_{\odot}$) model stars later in their evolutionary tracks,
closer to the giant phase. 
The basic features of our model is schematically depicted in fig.(\ref{fig2}), where, for future use, we indicate the position
of the Hydrogen burning shell by $x_1$, and $x_s$ is the position where the formula for the opacity switches 
inside the envelope, as these result from free-free transitions and electron scattering. Even this simple model
is rendered difficult to analyse in the presence of modified gravity. To begin with, there are a large number of 
algebraic and differential equations (a total of 45 such equations are listed in HS), and the presence of the two boundaries where
suitable boundary conditions need to be imposed, makes our task cumbersome. However, we carry out a detailed numerical 
analysis here and are able to obtain the basic physics of the effects of modified gravity inside the 
class of stellar objects that we consider. 

In order to study the specific model in the context of modified gravity, 
for a given $\Upsilon$, we start off with a suitably assumed value of
core temperature and the relative position in the envelope, where the mechanism of opacity changes. 
On doing so, we end up having a single parameter family of solutions for both the core and the envelope. 
These two parameters (let us say ${\mathcal A}$ for the core and ${\mathcal B}$ for the envelope)\footnote{${\mathcal A}$
is the quantity $\psi_c$ and ${\mathcal B}$ is $C_{kr}$ in the next section.}
are different of course. For different values of the envelope parameter ${\mathcal B}$, we integrate the dynamical 
equations of the envelope to obtain a family of envelope solutions. Similarly for different values of the core-parameter 
${\mathcal A}$, we integrate the dynamical equations of the core to obtain a family of core solutions. By fitting the two families 
of solutions (one on the core side and the other on the envelope side) at the core-envelope junction, we end up with a 
valid tuple of the parameters $({\mathcal A},{\mathcal B})$, generating a viable solution for the
entire star, respecting the continuity in stellar parameters like mass, pressure, and temperature.

From this complete solution, we can obtain the
isothermal core temperature as well as the relative position where the mode of
opacity changes inside the envelope. If these newly obtained values do not fall
within $1\%$ of the previously assumed values we started with, we have to iterate
the procedure, this time starting with these newly obtained values of core
temperature and the relative position. We continue this iterative process until
the solution converges. From the converged solution, we obtain all the stellar parameters for the given $\Upsilon$. Carrying this out for different $\Upsilon$ values, we end up with all the stellar parameters as a function of $\Upsilon$.

Having obtained the luminosity and the radius in terms of $\Upsilon$ in our model, we will need an 
independent estimate of the same in order to put possible bounds on the modified gravity parameter. 
For simplicity, if we assume a conservative estimate of a $3\%$ error margin on these observables,
then such an estimate can be obtained. We will see that this leads to a reasonably tight bound
on $\Upsilon$. 

\subsection{The modified gravity setup}
\label{modgrav}

To keep the discussion general at this stage, we will start by taking the Newtonian limit of the stress tensor inside
a stellar object, that in GR is given by $T^{\mu}_{\nu} = {\rm diag}(-\rho c^2, P_{rad}, P_{\perp}, P_{\perp})$ where
$c$ is the speed of light and we have allowed for the fact that in general an anisotropy might be present, with
$P_{rad}$ being the radial pressure and $P_{\perp}$ being the tangential one, and spherical symmetry dictating that
such tangential pressures along the non-radial directions should be equal. Now we consider a generic Friedman-Robertson-Walker
metric in the flat space-time limit, given by
\begin{equation}
ds^2 = -\left(1 + 2\Phi(r)\right)dt^2 + a(t)^2\left(1-2\Psi(r)\right)\left[dr^2 + r^2\left(d\theta^2 + \sin^2\theta d\phi^2\right)\right]~,
\end{equation}
where we will set the scale factor $a(t)$ to unity, since we are interested only in a static situation. Here, 
$\Phi(r)$ being the Newtonian potential, and we have $\Phi(r),\Psi(r) \ll 1$. That the energy momentum tensor is
covariantly conserved, i.e $D_{\mu}T^{\mu\nu}=0$ (with $D_{\mu}$ being the covariant derivative), then gives
in the Newtonian limit, 
\begin{equation}
\frac{dP_{rad}}{dr} = -\rho c^2\frac{d\Phi}{dr} +\frac{2}{r}\left(P_{\perp}-P_{rad}\right)\left(1-r\frac{d\Psi}{dr}\right)=0
\label{TOVA2}
\end{equation}
We then see that terms involving $\Psi$ can only come into the picture for theories with an anisotropy. 
In beyond-Horndeski theories,
the differential equations for the potential $\Phi(r)$ and $\Psi(r)$ were written down in \cite{SaksteinPRD} and read
\begin{equation}
\frac{d\Phi}{dr} = \frac{GM_r}{c^2r^2} + \frac{\Upsilon}{4}\frac{G}{c^2}\frac{d^2M_r}{dr^2}~,~~
\frac{d\Psi}{dr} = \frac{GM_r}{c^2r^2} - \frac{5\Upsilon}{4}\frac{G}{c^2 r}\frac{d M_r}{d r}~.
\label{phipsider}
\end{equation}
Here, $\Upsilon$, as mentioned before, is the parameter arising in the STT, 
with $G$ being the Newton's constant and $M_r$ denoting the mass up to radius $r$. Now, 
substituting eq.(\ref{phipsider}) in eq.(\ref{TOVA2}), and assuming an isotropic situation 
$P_{rad} = P_{\perp}=P$, we obtain the final form of the pressure balance equation
\begin{equation}
\frac{dP}{dr} = - \frac{GM_r\rho}{r^2} - \frac{\Upsilon}{4}G\rho\frac{d^2M_r}{dr^2}~,~~\frac{dM_{r}}{dr}=4\pi r^{2}\rho ~,
\label{TOVA3}
\end{equation}
with the second relation giving the mass in terms of the density.  Note that as pointed out in \cite{Saltas}, 
eq.(\ref{TOVA3}) can be recast into standard form, but with an effective Newton's constant $G_{eff}$, given as
\begin{equation}
\frac{G_{eff}}{G} = 1 + \frac{\Upsilon}{4}\frac{r^2}{M_r}\frac{d^2M_r}{dr^2}~.
\label{Geff}
\end{equation}
The second term in eq.(\ref{Geff}) determines the nature of modified gravity inside the star. Namely, from
\begin{equation}
\frac{d^2M_r}{dr^2} = 8\pi r\rho + 4\pi r^2\frac{d\rho}{d r}~,
\label{secondder}
\end{equation}
we can glean that since $d\rho/d r <0$ and for large $r$, the second term in eq.(\ref{secondder}) dominates over the
first, hence overall gravity weakens, at least for homogeneous low mass stars. However, close to $r=0$, if 
the fall off of the density is not too steep, the first term dominates, and gravity gets stronger (which was noted by \cite{Saito}). These are however
just ``rule of the thumb'' statements as pointed out in \cite{SaksteinRev}, and for the stellar model that we consider
here, a more detailed analysis is required. 

\subsection{The stellar equations}
\label{steq}

We will first write down the basic equations for the stellar interior that we consider, 
in the Newtonian limit of GR discussed above. We follow the notations of HS, and mention that 
the contents of this subsection can be found in textbooks, and we refer the reader to some of the 
excellent resources \cite{Schwarzschild},\cite{Chandrasekhar} - \cite{Kippenhahn}, 
In what follows, the subscripts have specific meanings. The subscript
$1$ on a variable stands for its value at the core envelope junction, $i$ stands for its value on the internal (core) side,
$e$ stands for its value on the envelope side, and $s$ stands for its value 
at the junction at which the opacity mechanism changes (see fig.(\ref{fig2})). 

Let us begin with the equation of state. In degenerate material (present in the core), in terms of the 
Fermi function defined as
\begin{equation}
F_{\nu}(\psi)=\int_{0}^{\infty} \frac{u^{\nu}}{e^{u-\psi}+1} du~,
\end{equation}
where $\psi$ is the degeneracy function \cite{HoyleSchwarzschild} (see also \cite{DeMarque}) and 
we have that the pressure $P$ and the density $\rho$ are given by
\begin{equation}
P=\frac{8\pi}{3h^{3}}(2mkT)^{3/2}kTF_{3/2}(\psi)~,~~
\rho=\frac{4\pi}{h^{3}}(2mkT)^{3/2}\mu_{i}HF_{1/2}(\psi)~,
\end{equation}
where $k$ is Boltzmann's constant, $\mu$ is the mean molecular weight, $h$ is Planck's constant, and
$m$ the electron mass. Note that HS ignores the ion pressure in the core. 
However, a later computation due to Hayashi \cite{Hayashi} shows that including this makes little
difference in the radius and luminosity of stars that have just left the main sequence, a scenario that
we are interested in here. 

In non-degenerate material, we assume the ideal gas relation
\begin{equation}
P=\frac{k}{\mu H}\rho T~,
\label{pnondeg}
\end{equation}
where $H$ is the mass of the Hydrogen atom. 
The hydrostatic equilibrium condition will play a central part in our analysis. 
In the Newtonian limit of GR, this is simply obtained by setting $\Upsilon = 0$ in eq.(\ref{TOVA3}). 

The thermal equilibrium conditions are then recorded. The core is assumed to be isothermal, at
temperature $T_1$. For radiative equilibrium in the envelope, we 
have\footnote{Note that the derivation of this equation follows from the pressure gradient of radiation. However,
as explained in \cite{SaksteinPRD}, this is not affected by gravity and hence does not change in the
theory of modified gravity that we consider.}
\begin{equation}
\frac{dT}{dr}=-\frac{3}{4ac}\frac{\kappa\rho}{T^{3}}\frac{L_r}{4\pi r^{2}}
\label{dTdr}
\end{equation}
where the opacities are
\begin{equation}
\kappa=3.68\times 10^{22}\left(1+X\right)\frac{\rho}{T^{3.5}}~,{\rm and}~~\kappa=0.19\left(1+X\right)
\label{op}
\end{equation}
for free-free scattering taking place in the outer part of the envelope, and for electron scattering taking place 
in the inner part, respectively, with $X$ being the Hydrogen fraction. Here, as discussed after eq.(\ref{Kramer}), 
we have, in Kramer's formula, set $(1-Z)\sim 1$.
Also note that in eq.(\ref{dTdr}), $a$ is the radiation constant, defined 
as $4\sigma/c$, with $\sigma$ being the Stefan-Boltzmann constant. 
For the isothermal core with a temperature $T$, we further have $\frac{dT}{dr}=0$.
Next we come to nuclear-energy production. With $R$ the radius of the star, its luminosity is 
\begin{equation}
L=4\pi \int_{0}^{R}\epsilon\rho r^{2}dr
\label{Lum1}
\end{equation}
where $\epsilon=\epsilon_{CN}  + \epsilon_{pp}$, where $\epsilon$ denotes the amount of energy released per unit 
mass per unit time and the subscripts denote the values for the Carbon-Nitrogen cycle and the $p-p$ chain
reactions, respectively. Now, we will take
\begin{equation}
\epsilon_{CN}=600\rho X \frac{X_{CN}}{0.01}\left(\frac{T}{20\times10^6}\right)^{15}~,~~
\epsilon_{pp} = 0.5\rho X^2\left(\frac{T}{15\times 10^6}\right)^4~.
\label{Lum2}
\end{equation}
While the second equation is standard, the first requires some careful explanation. 
We note that the exponent $15$ appearing in eq.(\ref{Lum2}) was {\it assumed} by Hoyle and Schwarzschild 
and differs from the more conventional exponent close to $20$ for the C-N cycle. This exponent of 
$15$ is more appropriate for giant stars, for which the temperature at the thin hydrogen burning shell
is more than $20$ million Kelvins. In our case, although the star is in the sub-giant phase, 
the choice of a negligible shell thickness (as discussed in subsection \ref{model}) makes it more appropriate 
to consider the lower exponent in eq.(\ref{Lum2}). That this gives reliable estimates of the 
temperature, luminosity etc. for a $1.1M_{\odot}$ star at its turnoff point was noted by HS. 

Now, we note that in eq.(\ref{Lum2}), along with the values given in eq.(\ref{vals}) below, 
the C-N cycle starts to dominate the energy generation process
at shell temperatures more than $15.2$ million Kelvins. As we will see later, the shell temperature ranges that we 
will be interested in here will range from $17.5$ to $18.1$ million Kelvins, for which this will be true. 
In fact, for a shell temperature of $17.5 \times 10^6$ K, in the model considered here, the energy generation
by the C-N cycle is roughly $5$ times that by the $p-p$ chain, and increases to $7$ times that of the
$p-p$ chain for a shell temperature of $18.1 \times 10^6$ K. Therefore, as a convenient mathematical
simplification, we will ignore the $\epsilon_{pp}$ here, as we had mentioned in section \ref{model}.  

Next, we come to the homology invariants (for a recent discussion on the use of the homology invariants
in stellar physics, see e.g \cite{Kippenhahn}), defined from 
\begin{equation}
U=\frac{d \ln M_r}{d \ln r}~,~~V=-\frac{d \ln P}{d \ln r}~,~~n+1=\frac{d \ln P}{d \ln T}=
\begin{cases}
\frac{16\pi acG}{3}\frac{T^4}{\kappa P}\frac{M_{r}}{L},~\text{in the envelope}
\\
\infty,~\text{in the isothermal core}~.
\end{cases}
\label{hom}
\end{equation}
Finally, we record the constants assumed. These are 
\begin{equation}
X=0.9~,~X_{CN}=0.0005~,~\mu_{e}=0.533~,~\mu_{i}=1.333~.
\label{vals}
\end{equation}
These are the most important equations that govern the physics of the star that we consider, in the 
Newtonian limit of GR. We now move over to the changes in these, in the presence of modified gravity. 

\subsection{The transformed stellar equations in modified gravity}
\label{Transformed}

The change in the hydrostatic equilibrium condition as given in eq.(\ref{TOVA3}) modifies the entire analysis
non-trivially in the context of the beyond Horndeski theory. In this subsection, we will write down the modified 
equations that follow. We first define the non-dimensional variable
\begin{equation}
\xi = \frac{4\pi\mu_i H}{h^{3/2}}\left(2mkT_1\right)^{1/4}\left(2mG\right)^{1/2}r
\end{equation}
In terms of $\xi$, the modified hydrostatic equilibrium equation in the core reads
\begin{equation}
\frac{1}{\xi^{2}}\frac{d}{d\xi}\Big(\xi^{2}\frac{d\psi}{d\xi}\Big)=
-\Big[F_{1/2}(\psi)\Big(1+\frac{3\Upsilon}{2}\Big)+\frac{3\Upsilon}{2}\xi\frac{d}{d\xi}F_{1/2}(\psi)
+\frac{\Upsilon}{4}\xi^{2}\frac{d^{2}}{d\xi^{2}}F_{1/2}(\psi)\Big]
\label{mhec}
\end{equation}
with boundary conditions $\psi$=$\psi_{c}$, $\frac{d\psi}{d\xi}=0$ at $\xi=0$. 
Note that the quantity $\psi$ now depends on $\Upsilon$ and hence affects all the 
core variables listed in the previous section, either explicitly or implicitly. 

For example, we have in the core, 
\begin{equation}
M_{r}=\frac{h^{3/2}}{4\pi \mu_{i}^{2}H^{2}}(2mkT_{1})^{3/4}(2mG)^{-3/2}\bar{\phi}~,~~
\bar{\phi}=\int_{0}^{\xi}F_{1/2}(\psi){\bar{\xi}}^{2}d\bar{\xi}~,
\label{mass}
\end{equation}
where, to make the notation more compact, we have defined the quantity $\bar{\phi}$.
The other core variables are also summarized as
\begin{eqnarray}
T=T_{1}~,~~U&=&\frac{\xi^3 F_{1/2}(\psi)}{\bar{\phi}}~,~~
V=\frac{3}{2}\frac{F_{1/2}(\psi)}{F_{3/2}(\psi)}\frac{\bar{\phi}}{\xi}+\frac{3\Upsilon}{8}\xi
\frac{F_{1/2}(\psi)}{F_{3/2}(\psi)}\frac{d^{2}\bar{\phi}}{d\xi^{2}}
\end{eqnarray}
Next, we come to the modified equilibrium equations for the envelope. 
With non-dimensional variables $p$, $q$, $t$, and $x$ defined from
\begin{equation}
P=p\frac{GM^{2}}{4\pi R^{4}}~,~T=t\frac{\mu_e H}{k}\frac{GM}{R}~,~M_{r}=qM~,~r=xR~,
\end{equation}
these can be shown to be given by 
\begin{equation}
\frac{dp}{dx}=-\frac{q}{x^{2}}\frac{p}{t}-\frac{\Upsilon}{4}\frac{p}{t}\frac{d^{2}q}{dx^{2}}~,~
\frac{dq}{dx}=\frac{px^{2}}{t}~,~\frac{dt}{dx}=
\begin{cases}
-C_{kr}\frac{p^{2}}{x^{2}t^{8.5}},  \text{$x>x_{s}$}
\\
-C_{el}\frac{p}{x^{2}t^{4}},  \text{$x<x_{s}$}
\end{cases}
\label{env1}
\end{equation}
with the boundary conditions $q=1,\hspace{0.2cm}t=0,\hspace{0.2cm}p=0$ at $x=1$,
and we have also defined
\begin{equation}
C_{kr}=\frac{3}{4ac}\frac{3.68\times 10^{22}(1+X)}{(4\pi)^{3}}\Big(\frac{k}{\mu_{e}HG}\Big)^{7.5}\frac{LR^{0.5}}{M^{5.5}}~,~
C_{el}=\frac{3}{4ac}\frac{0.19(1+X)}{(4\pi)^{2}}\Big(\frac{k}{\mu_{e}HG}\Big)^{4}\frac{L}{M^{3}}~.
\end{equation}
These satisfy the relation
\begin{equation}
C_{el}=C_{kr}\frac{p_s}{t_s^{4.5}}~,
\end{equation}
which arises due to the continuity of $\frac{dt}{dx}$ at $x_{s}$. Also, the homology invariants are defined
in the envelope as
\begin{equation}
U=\frac{px^{3}}{qt}~,~
V=\frac{q}{xt}+\frac{\Upsilon}{4}\frac{x}{t}\frac{d^{2}q}{dx^{2}}~,~
n+1=\begin{cases}
\frac{q}{C_{kr}}\frac{t^{8.5}}{p^2}, \text{ for $x>x_s$}
\\
\frac{q}{C_{el}}\frac{t^4}{p}, \text{ for $x<x_s$}
\end{cases}
\label{hommod}
\end{equation}

We will now specify the matching conditions between the core and the envelope. 
\begin{itemize}
\item As is clear from eq.(\ref{hom}) along with eq.(\ref{vals}), matching of homology invariants at $x_1$ requires
\begin{equation}
\frac{U_{1i}}{U_{1e}}=\frac{\mu_{1i}}{\mu_{1e}}=\frac{V_{1i}}{V_{1e}}=2.5~,~
(n+1)_{1e} ~ \text{finite}
\label{fitting}
\end{equation}
\item Fitting of $M_{r}$ at $x_1$ requires that 
$M_{r1i}=q_{1e}M$
\item Fitting of $r$ at $x_1$ requires that $r_{1i}=x_{1e}R$
\item Fitting of opacity at $x_s$ requires that $L$ as computed from $C_{kr}$ must equal $L$ 
as computed from $C_{el}$.
\end{itemize}
	
Finally, we record the expressions for the energy production in the thin shell at $x_1$. Writing the
pressure balance equation of eq.(\ref{TOVA3}) in terms of the the homology invariants
in eq.(\ref{hom}), we obtain in the thin shell approximation, 
\begin{equation}
P\simeq P_{1}\left(\frac{r}{r_1}\right)^{-V_{1e}}~,~T\simeq T_{1}\left(\frac{r}{r_1}\right)^{-V_{1e}/(n+1)_{1e}}~,
\end{equation}
where now $V_{1e}$ is $\Upsilon$ dependent, and given by the second relation in eq.(\ref{hommod}). 
Substituting these results in eq. (\ref{Lum2}), and using eq.(\ref{pnondeg}), we get $L=L_{CN}$, with
\begin{eqnarray}
L_{CN}\simeq 600X\frac{X_{CN}}{0.01}\rho_{1e}^{2}\Big(\frac{T_1}{20 \times 10^{6}}\Big)^{15}
\frac{4\pi r_{1}^{3}}{V_{1e}[2+13/(n+1)_{1e}]-3}.
\label{LuminF}
\end{eqnarray}
Of course, putting $\Upsilon = 0$ in any of the above equations will give the ordinary Newtonian equations and results.

\section{Numerical Analysis and Results}

We now present the detailed numerical analysis of the modified equations, discussed in the last 
subsection.\footnote{All numerical analysis was carried out at the High Performance Computing (HPC) facility
at IIT, Kanpur, India and we have used C++ codes, along with appropriate Mathematica subroutines. }
The reader would by now have realized the relevance of our qualitative discussion in the introduction. 
There, we schematically illustrated the procedure by two quantities that we called ${\mathcal A}$ and ${\mathcal B}$. Clearly,
identifying ${\mathcal A}$ as $\psi_c$ and ${\mathcal B}$ as $C_{kr}$ brings us to the relevant discussion in the 
context of the model described in the last section. 

\subsection{Numerical Analysis}
\label{Numerical}

From the equations presented there, we see that for every choice of $M$ and $q_1$, if $T_1$ (core temperature) 
and $x_s$ (the relative position in the envelope where the opacity changes from being due to free-free
transitions to free-electron scattering) are provided from beforehand, then the
solution to the envelope equations will be solely dependent upon the choice of the
value of $C_{kr}$. Similarly the solutions to the core equations will be solely
governed by the choice of the initial value $\psi_{c}$. Thus $C_{kr}$ serves as
the single parameter for the family of envelope solutions and $\psi_{c}$ serves as
the single parameter for the family of core solutions.

The position of the junction ($x_1$) from the envelope side is achieved when $q(x)$
reaches value $q_1$. Similarly the position of the junction
($\xi_1$ for each $\psi_{c}$) from the core side is obtained when the mass of the core
$(M_{r})$ attains value $q_1M$.
To get the valid solution for the entire stellar object (i.e. core solution
merging continuously with the envelope solution at the core-envelope junction) we
need the fitting conditions via the homology invariants discussed before.

For an initially chosen $T_1$, we get the values of tuple ($U_{1i},V_{1i}$) corresponding to different values 
of the free parameter $\psi_{c}$. We plot the corresponding curve (calling it C-curve, where C stands for core) in a 
$U-V$ plane. Now for an initially chosen $x_s$, we get the values of tuple ($U_{1e},V_{1e}$) corresponding to 
different values of the free parameter $C_{kr}$. After making an appropriate jump by a factor of $2.5$ according to 
the fitting condition of eq.(\ref{fitting}), we plot the corresponding curve (calling it E-curve, where E stands for envelope) 
in the same $U-V$ plane. From such a U-V plane we get the values of $C_{kr}$ and $\psi_{c}$ corresponding to 
the intersecting point of the C-curve and the E-curve, which will yield perfect 
fit\footnote{The fitting is done within a tolerance limit of $\pm0.001$} 
of the two solutions (core and envelope) for the chosen $T_1$ and $x_s$. Now corresponding to these values of 
$T_1$, $x_s$, $C_{kr}$, $\psi_{c}$, we obtain the stellar radius $R$ (see discussion after eq.(\ref{fitting})).

Further, we obtain the luminosity $L$ from the expression of $C_{kr}$. Now from the  
expression of the luminosity in the shell, we obtain a polynomial in $T_1$ of degree 15 (see 
eq.(\ref{LuminF})). We solve this to get $T_1$. From the requirement of continuity of opacity in
the envelope, we get $x_s$ (see discussion after eq.(\ref{fitting})). If these newly obtained values of $T_1$ and $x_s$ are
within $1\%$ of the initially assumed values of $T_1$ and $x_s$, then we stop the
iteration, else we continue the aforementioned procedure until the condition of convergence
gets satisfied, but at every subsequent iteration, we start with the obtained values of $T_1$
and $x_s$ from the immediately previous iteration. 

According to HS, the initial values of $T_1$ and $x_s$ are 
to be chosen {\it shrewdly}. This is crucial in our analysis and we have adhered to their suggestion.
First we take a particular value of $T_1$, typical of the class of stars we are considering, and then obtain the 
C-curve in the $U-V$ plane. We then choose $x_s$ accordingly, such that the corresponding E-curve 
has an intersection with the aforementioned C-curve in the $U-V$ plane. We obtain the stellar parameter values for 
different $\Upsilon$ values, as tabulated in a while.

There are a few computational subtleties that are best mentioned at this point. When we merge the 
expressions appearing in eq.(\ref{env1}), we obtain
\begin{equation}
\frac{dp}{dx}=-
\begin{cases}
\frac{\Big(\frac{q}{x^2}pt +\frac{\Upsilon}{2}xp^{2}+\frac{C_{kr}}{4}\Upsilon\frac{p^4}{t^{9.5}}\Big)}
{\Big(t^{2}+\frac{\Upsilon}{4}x^{2}p\Big)}, \text{ for $x>x_s$}
\\
\frac{\Big(\frac{q}{x^2}pt +\frac{\Upsilon}{2}xp^{2} +\frac{C_{el}}{4}\Upsilon\frac{p^3}{t^{5}}\Big)}{\Big(t^{2}
	+\frac{\Upsilon}{4}x^{2}p\Big)}, \text{ for $x<x_s$}
\end{cases}
\end{equation}
Now, considering the boundary condition $q=1, t=0, p=0$ at $x=1$, 
the third term in the numerator of this equation blows up, when we take very small numerical values, say 
$t=10^{-7},\hspace{0.2cm}p=10^{-7}\hspace{0.2cm}$ at $x=1.0$, to start with. Initially near the surface 
$x=1.0$, the third term turns out to be the most dominant term, in presence of $\Upsilon$. Thus
\begin{equation}
\frac{dp}{dx}\simeq-\frac{\Big(
	\frac{C_{kr}}{4}\Upsilon\frac{p^4}{t^{9.5}}\Big)}{\Big(t^{2}+\frac{\Upsilon}{4}x^{2}p\Big)}\simeq -\frac{C_{kr}}{x^2}\frac{p^3}{t^{9.5}}
\label{evo1}
\end{equation}
Without $\Upsilon$, only the first term remains, which is thus the dominating term for GR, i.e
\begin{equation}
\frac{dp}{dx}=-\frac{qp}{x^2t}
\end{equation}
Hence the initial evolution of the equations from $x=1.0$ to $x=(1.0 - 10^{-6})$ say, is distinctly different for the 
two cases, i.e one with non zero $\Upsilon$ and the other with $\Upsilon=0$. The values of $p$, $q$, $t$ are different 
at $x=(1 - 10^{-6})$ for the two cases, although they start  from the same value at $x=1.0$. Thus, once we 
evolve along the STT path, we cannot retrace back the GR results by simply putting $\Upsilon=0$. 
Hence we have chosen to evolve our equations initially via the GR path (from $x=1.0$ to $x=(1.0 - 10^{-6})$) and then 
evolved them using eq.(\ref{evo1}), because after the first step, the parameters take up finite 
values, which no longer poses the blowing-up issue. Using this technique we have a handle of getting GR 
results, by putting $\Upsilon=0$ in all the featured equations in this paper.

\subsection{Results and discussions}
\label{Results}

We are now ready to present our main results. First, we
present the results on our numerical analysis based on the equations of 
section \ref{Transformed} and the methods of section \ref{Numerical}, and discuss these. 
Then, we will discuss a possible bound on $\Upsilon$. 

For M$=1.1M_{\odot}$ and $q_{1}=0.10$ we perform the above numerical computation for different values of $\Upsilon$ 
and obtain a Table (\ref{Table1}) of different physical as well as non-dimensional variables.
\begin{table}[h!]
\begin{center}
\begin{tabular}{|c|c|c|c|c|c|c|}
\hline 
& \multicolumn{6}{c|}{$\Upsilon$} \\ 
\cline{2-7}
& -0.34 & -0.24 & -0.14 & 0.0 & 0.14 & 0.24 \\ 
\hline 
$T_{1}\times(10^{-6})$ & 18.08 & 18.05 & 18.00 & 17.86 & 17.65 & 17.45 \\ 
\hline 
$C_{kr}\times(10^{6})$ & 2.90 & 2.79 & 2.67 & 2.52 & 2.32 & 2.19 \\ 
\hline 
$\psi_c$ & -0.65 & -0.11 & 0.50 & 1.44 & 2.41 & 3.15 \\ 
\hline 
log$x_s$ & -0.51 & -0.54 & -0.56 & -0.61 & -0.63 & -0.66 \\ 
\hline 
log$p_s$ & 1.23 & 1.28 & 1.34 & 1.43 & 1.49 & 1.56 \\ 
\hline 
log$q_s$ & -0.26 & -0.27 & -0.28 & -0.30 & -0.31 & -0.33 \\ 
\hline 
log$t_s$ & -0.27 & -0.25 & -0.24 & -0.22 & -0.21 & -0.19 \\ 
\hline 
log$x_1$ & -0.99 & -1.04 & -1.10 & -1.17 & -1.23 & -1.26 \\ 
\hline 
log$p_1$ & 1.90 & 2.06 & 2.21 & 2.41 & 2.57 & 2.68 \\ 
\hline 
log$t_1$ & 0.03 & 0.07 & 0.10 & 0.15 & 0.18 & 0.20 \\ 
\hline 
$V_{1e}$ & 0.84 & 0.93 & 1.01 & 1.11 & 1.19 & 1.24 \\ 
\hline 
$U_{1e}$ & 0.76 & 0.68 & 0.61 & 0.54 & 0.49 & 0.47 \\ 
\hline 
$(n+1)_{1e}$ & 1.95 & 2.09 & 2.21 & 2.36 & 2.47 & 2.53 \\ 
\hline 
$r_{1}\times(10^{-10})$ & 0.63 & 0.58 & 0.53 & 0.47 & 0.42 & 0.39 \\ 
\hline 
$10\times$log$(\frac{R}{R_{\odot}})$ & -0.54 & -0.32 & -0.15 & 0.04 & 0.13 & 0.17 \\ 
\hline 
$10\times$log$(\frac{L}{L_{\odot}})$ & 5.31 & 5.03 & 4.76 & 4.41 & 4.00 & 3.75 \\ 
\hline 
$\log(\rho_{1e})$ & 1.74 & 1.80 & 1.86 & 1.96 & 2.06 & 2.14 \\ 
\hline
$\log(T_{1e})$ & 7.22 & 7.23 & 7.25 & 7.28 & 7.30 & 7.31 \\
\hline
log($T_{eff}$) & 3.92 & 3.90 & 3.89 & 3.87 & 3.85 & 3.84 \\
\hline 
\end{tabular}
\caption{Dependence on $\Upsilon$ of the various physical parameters listed. ($M=1.1M_{\odot}$, $q_{1}=0.10$)}
\label{Table1}
\end{center}
\end{table}
These results are summarised as follows. 
\begin{itemize}
\item The radius $R$ of the star, the density $\rho_{1e}$, and the temperature on the envelope side $T_{1e}$ of 
the nuclear burning shell increase with increase in $\Upsilon$.
\item The size of the core $r_1$, the core temperature $T_1$, the overall luminosity $L$ of the star and
the effective temperature $T_{eff}$ at the surface decrease with increase in $\Upsilon$.
\end{itemize}
As a result of the above, with increasing $\Upsilon$, the span of the envelope increases, and the radial position $(x_s)$, where 
the switch of opacity occurs, comes closer to the core. Clearly, these point to very non-trivial physics in the stellar interior, as 
we have mentioned before. The decrease in the luminosity $L$ and the effective temperature $T_{eff}$ is consistent
with the fact that overall gravity weakens inside the stellar object (its radius increases), and it becomes dimmer and cooler. 
However, we note that near the core, the opposite happens. The core radius decreases appreciably as one
increases $\Upsilon$, indicating that gravity is stronger near the core than what would happen in the $\Upsilon = 0$
case. In addition, the core temperature also decreases, although by a small amount. To balance the strong
gravity, the density at the core increases by a large amount as one increases $\Upsilon$. 
We have checked that this is not an artefact of our choice of $M=1.1M_{\odot}$. The same trend is followed by
models with $M = 1.15$ and $1.2 M_{\odot}$, although, as mentioned before, these masses are more appropriate
for modeling stars near the giant phase. 

Now, from Table \ref{Table1}, we obtain a relationship between $L$ and $R$ as a function of 
$\Upsilon$. We see a monotonic decrease in $\log R/R_{\odot}$
and a monotonic increase in $\log L/L_\odot$ with decrease in $\Upsilon$. 
\begin{figure}[h]
\centering
\begin{minipage}{.5\textwidth}
\centering
\includegraphics[width=.9\linewidth]{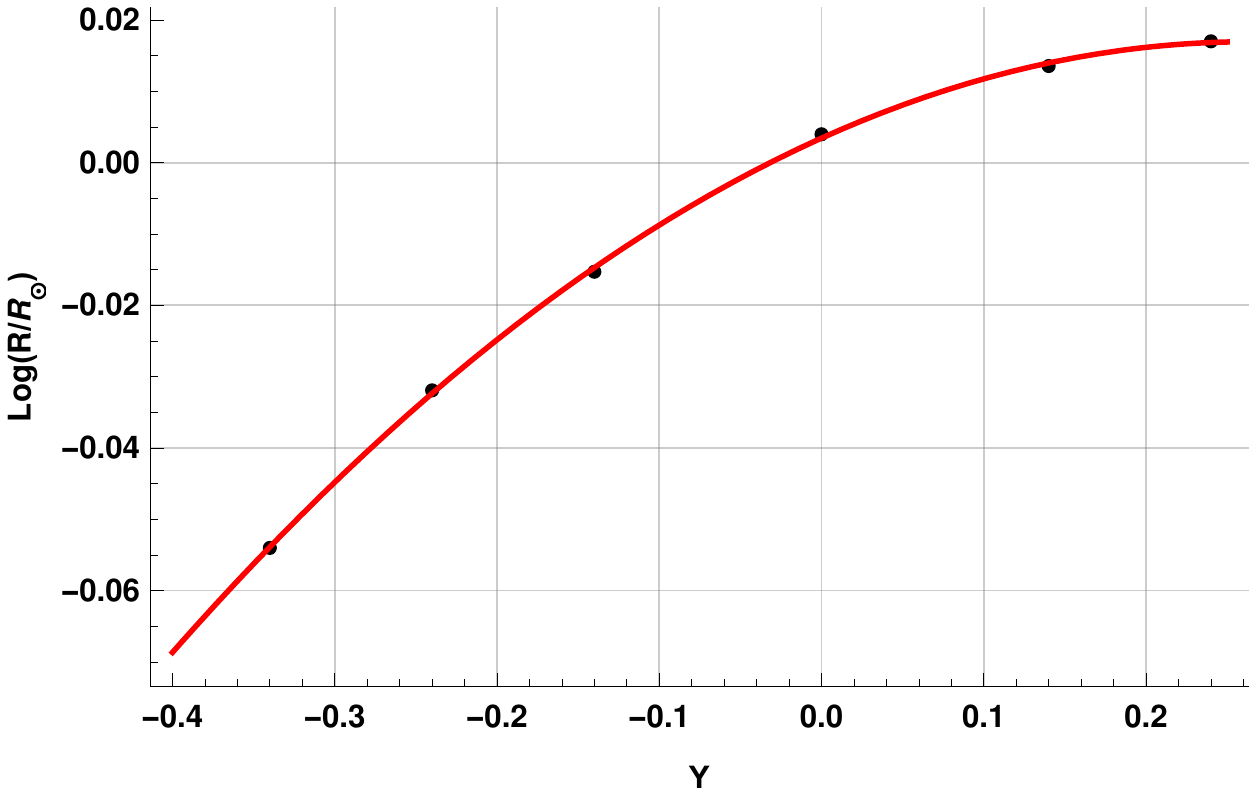}
\caption{$\log R/R_{\odot}$ vs $\Upsilon$}
\label{figR}
\end{minipage}%
\begin{minipage}{.5\textwidth}
\centering
\includegraphics[width=.9\linewidth]{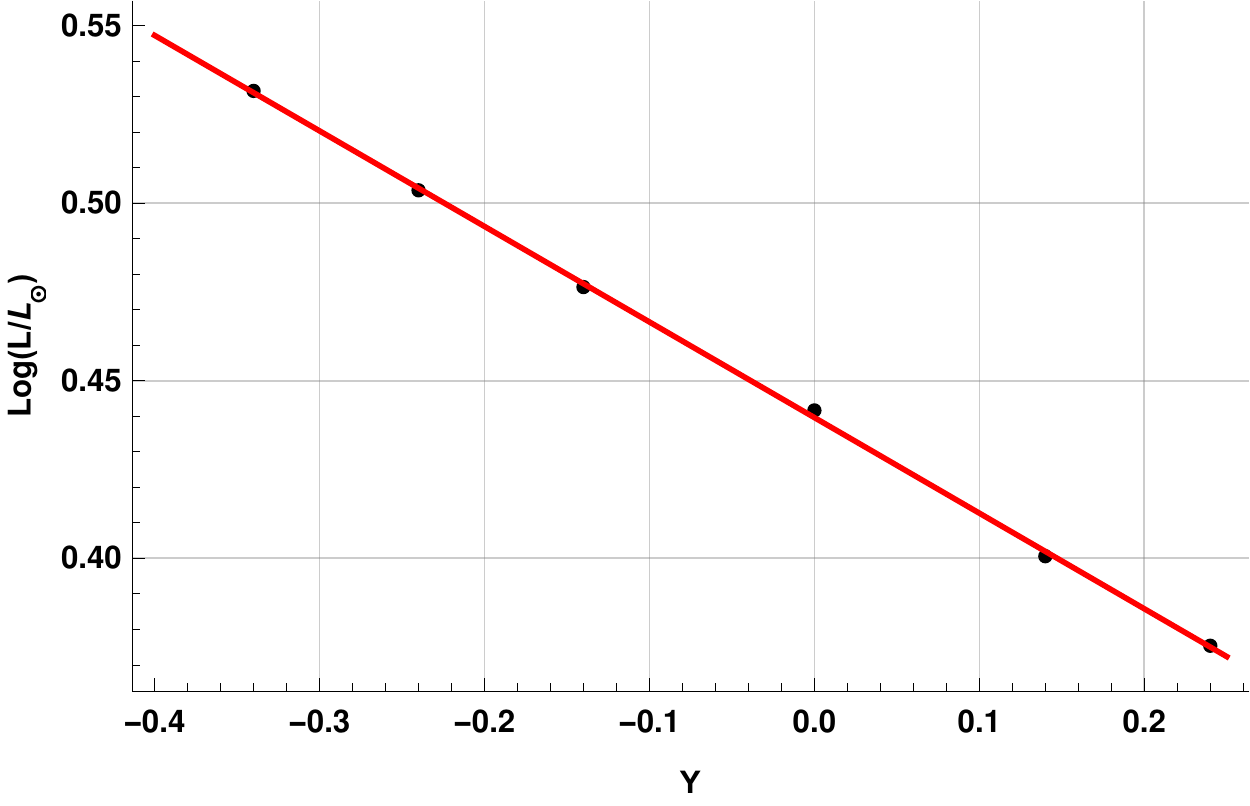}
\caption{$\log L/L_{\odot}$ vs $\Upsilon$}
\label{figL}
\end{minipage}
\end{figure}
In figs.(\ref{figR}) and (\ref{figL}), we plot the obtained value of the radius and the luminosity as
a function of $\Upsilon$. These can be fitted as 
\begin{equation}
\log\left(\frac{R}{R_{\odot}}\right) = 0.003 + 0.103\Upsilon - 0.194\Upsilon^{2}~,~~
\log\left(\frac{L}{L_{\odot}}\right)=0.440 - 0.269\Upsilon~,
\label{LRFit}
\end{equation}
and we have plotted the actual results given by the dots, along with the fitted curve. 
One would now ideally like to use a mass-radius or a mass-luminosity relation for pop-II stars 
to get an estimate for $\Upsilon$ from eq.(\ref{LRFit}). Unfortunately, while such relations are
available in the literature (see, e.g., \cite{Eker} for a recent work, which also gives a nice
historical account of the development of the subject in its introductory section) for solar
neighborhood pop-I stars, these are not very well formulated for pop-II stars as of now. 

We can however envisage a bound on $\Upsilon$ from possible error estimates in the 
measurement of the luminosity. For example, if we assume a very conservative
$\sim 3\%$ error in $L$ (for $\Upsilon = 0$), it is checked from the second relation 
of eq.(\ref{LRFit}) that 
\begin{equation}
-0.05 < \Upsilon < 0.04~.
\label{FinalRange}
\end{equation}
Of course, it has to be checked that such a range is consistent with the errors in the numerical
analysis that we have carried out, namely that the numerical error in computing the 
luminosity is less than $\sim 3\%$. That this is the case is easily checked. 
One might ask if the conclusions above change if the model here
was modified to include a small metallicity. This is an important source of degeneracy as pointed out
by Koyama and Sakstein \cite{SaksteinPRD}. 
Note that in this model, the only places where an 
explicit dependence on metallicity will show up are in the mean molecular weights of the core and the 
envelope, and in Kramer's opacity formula (through a multiplicative factor of $(1-Z)$).
Including, say, $Z = 0.001$ in our analysis results in negligible changes in the luminosity and the effective
temperature at $\Upsilon = 0$ (the change is $\sim 0.5\%$ for $L$, while for $T_{eff}$ it is even lesser). 
Hence, keeping the theory as GR and changing the 
metallicity in this model should not alter the conclusion of eq.(\ref{FinalRange}). 

\section{Conclusions}

It is well known that modified gravity theories beyond Horndeski, which are ghost free, typically exhibit a 
partial breaking of the Vainshtein mechanism inside stellar objects. This leads to a novel modification of 
gravity inside such objects, and in the Newtonian limit leads to an alteration of the pressure balance equation
therein. This feature provides an exciting mechanism to test a class of modified gravity theories, using astrophysical
signatures. Whereas many of the works in this direction concentrated on stellar objects that satisfy a polytropic
equation of state (for example in dwarf stars), here we have taken the first step towards studying the effects of 
modifications of gravity inside a stellar object that has a core-envelope structure. In such a situation, the effects of 
modified gravity have to be matched at appropriate junctions, which involves a detailed and involved numerical analysis. 
The main results of this paper are contained in section \ref{Results}, in Table \ref{Table1} (and the discussions thereafter) 
and eq.(\ref{FinalRange}). 

For completeness, in Table(\ref{Table2}), we provide the currently available bounds on $\Upsilon$ ($\sigma$ denotes
standard deviation here, wherever available). 
\begin{table}[h!]
\begin{center}
\begin{tabular}{|c|c|c|}
\hline 
Reference & Lower Bound & Upper Bound  \\ 
\hline 
\cite{Saito} & $-0.67$ &  -\\ 
\hline
\cite{Sakstein} & - & $1.6$ \\
\hline
\cite{Jain} & $-0.22$ ($2\sigma$) & $0.27$ ($2\sigma$)\\ 
\hline 
\cite{Babichev} & $-0.44$ & - \\
\hline
\cite{Saltas} & - & $0.14$ \\
\hline
\cite{Saltas2} & $-1.8\times 10^{-3}$ ($2\sigma$) & $1.2\times 10^{-3}$ ($2\sigma$) \\
\hline
\cite{Tapo2} & $0$ & $0.47$ \\
\hline
\end{tabular} 
\caption{Various bounds on $\Upsilon$.}
\label{Table2}
\end{center}
\end{table}
Most of the bounds above have been obtained in the context of dwarf stars, excepting the one in \cite{Saltas2},
where a precision constraint using helioseismology was obtained. We have, in this paper,
constructed the bound of eq.(\ref{FinalRange}) from an analysis of the internal structure of pop-II stars in globular clusters. One can
see that even within the limitations of the model considered here, our bound is a good improvement from
the ones available till now, from stellar structure constraints. 

\begin{center}
{\bf Acknowledgements}
\end{center}
SC acknowledges discussions with A. Chatterjee and S. Sadhukhan on computational issues.

\end{document}